\documentclass{ws-ijmpd}
\usepackage[super,compress]{cite}
\usepackage{amsmath,amsfonts,amssymb}
\usepackage{color,verbatim,graphicx,float}

\begin{document}

\markboth{Jakub Mielczarek, Linda Linsefors and Aurelien Barrau}
{Silent initial conditions for cosmological perturbations with a change of space-time signature}

\newcommand{\dR}{\mathbb R}
\newcommand{\dC}{\mathbb C}
\newcommand{\dZ}{\mathbb Z}
\newcommand{\id}{\mathbb I}
\newcommand{\dT}{\mathbb T}
\newcommand{\red}[1]{\textcolor{red}{#1}}
\definecolor{Green}{rgb}{0,0.7,0}
\newcommand{\green}[1]{\textcolor{Green}{#1}}
\newcommand\be{\begin{equation}}
\newcommand\ee{\end{equation}}
\newcommand\h{\mathcal{H}}
\newcommand\p{\mathcal{P}}
\newcommand\g{\mathcal{G}}

\title{Silent initial conditions for cosmological perturbations with a change of space-time signature}

\author{Jakub Mielczarek$^{1}$, Linda Linsefors$^{2}$ and Aurelien Barrau$^{2}$}

\address{$1$ Institute of Physics, Jagiellonian University, {\L}ojasiewicza 11, 30-348 Cracow, Poland \\
$2$ Laboratoire de Physique Subatomique et de Cosmologie, UJF, INPG, CNRS, IN2P3
53, avenue des Martyrs, 38026 Grenoble cedex, France}

\maketitle

\begin{abstract}
Recent calculations in loop quantum cosmology suggest that a transition from a Lorentzian to 
an Euclidean space-time might take place in the very early Universe. The transition point leads 
to a state of silence, characterized by a vanishing speed of light. This behavior can be interpreted 
as a decoupling of different space points, similar to the one characterizing the BKL phase. 

In this study, we address the issue of imposing initial conditions for the cosmological 
perturbations at the transition point between the Lorentzian and Euclidean phases. 
Motivated by the decoupling of  space points, initial conditions characterized by  
a lack of correlations are investigated. We show that the ``white noise''  gains some  
support from analysis of the vacuum state in the deep Euclidean regime.
 
Furthermore, the possibility of imposing the silent initial conditions at the trans-Planckian
surface, characterized by a vanishing speed for the propagation of modes with wavelengths  
of the order of the Planck length, is studied. Such initial conditions might result from a 
loop-deformations of the Poincar\'e algebra. The conversion of the silent initial power spectrum 
to a scale-invariant one is also examined. 

\end{abstract}

\ccode{PACS numbers: 98.80.Qc}

\section{Introduction}

A key issue in constructing any model describing the evolution of the Universe is the 
initial value problem. At the classical level, the Cauchy initial conditions have to be clearly specified. 
Imposed, \emph{e.g.}, at the present epoch they allow for a (at least partial) reconstruction of the cosmic dynamics backward and forward 
in time. When dealing with the early Universe, the initial conditions may be also introduced \emph{a priori} in the distant past. 
In this case, the consequences of a given assumption about the evolution 
of the Universe can be studied. In particular, the impact of the initial conditions on the 
properties of the inflationary stage has often been addressed. Many questions about why a given inflationary 
trajectory rather than another is realized in Nature remain open and the debate about the ``naturalness" of inflation is going on.  
It is however widely believed that the answer should come from a detailed understanding of the pre-inflationary 
quantum era. One may hope that by taking into account the quantum aspects of gravity, 
some specific initial stages, leading to the proper inflationary evolution, will be naturally distinguished. 
An extreme example of such a state is given by the so-called Hartle-Hawking no-boundary 
proposal, which basically circumvent the problem of the initial conditions \cite{Hartle:1983ai}. 

Recent results in loop quantum cosmology (LQC) provide a new opportunity to address the problem
of initial conditions. Namely, it was shown that the signature of space-time can effectively change 
from Lorentzian to Euclidean at extremely high curvatures \cite{Cailleteau:2011kr,Bojowald:2011aa,
Mielczarek:2012pf}. This effect is basically due to the requirement of anomaly freedom, that 
is to the necessity to have a closed algebra of quantum-corrected effective constraints when holonomy corrections from loop quantum gravity are included (the situation is less clear when inverse-triad terms are also added \cite{Cailleteau:2013}).
The beginning of the Lorentzian phase seems to be a natural place where initial conditions 
should be imposed. In this study, we follow this idea and investigate a possible choice of initial 
conditions on --or around-- this surface. We address this issue in the case of cosmological perturbations.  

In the considered model, the transition between the Lorentzian stage and the Euclidean stage is 
taking place at the energy density (see, {\it e.g.} \cite{consist})
\begin{equation}
\rho= \frac{\rho_c}{2}, 
\end{equation}
where $\rho_{c}$ is the maximal allowed value of the energy density.  In loop quantum 
cosmology, the value of $\rho_{c}$, reached at the bounce, is usually assumed to 
be given by the area gap of the area operator in loop quantum gravity, which gives 
$\rho_{c} \sim \rho_{\text{Pl}}$,  where $\rho_{\text{Pl}}$ is the Planck energy density.  
Latest studies suggest that $\rho_{c} = 0.41 \rho_{\text{Pl}}$ but they require some extra 
assumptions. To remain quite generic, we assume in the following that $\rho_{c}=\rho_{\text{Pl}}$, 
that is the naturally expected scale. Because of this, the numerical values obtained in this 
paper should be considered as orders of magnitudes rather than accurate results but 
the main conclusions do not depend on this. 

The background geometry is assumed to be described by the 
Friedmann-Lemaitre-Robertson-Walker metric. In effective LQC, the dynamics of the scale factor $a$ is governed 
by the modified Friedmann equation (see \cite{lqc_review} for introductory reviews):
\begin{equation}
H^2 = \frac{8\pi G}{3} \rho \left(1- \frac{\rho}{\rho_c}\right).
\end{equation}
The equation is derived for the flat FRW case and the form of quantum corrections 
has been justified mostly for the kinetic energy domination regime of the scalar field.
In this article, we fix the value of the scale factor to $a=1$ at the beginning of the Lorentzian 
stage (that is when $\rho=\rho_{\text{c}}/2$).  
It is easy to show that for $\rho= \frac{\rho_{\text{c}}}{2}$, the value of the Hubble parameter 
is maximal and equal to 
\begin{equation}
H_{\text{max}} = \sqrt{\frac{8\pi G \rho_c}{12}}. 
\end{equation} 

The equation of motion for  scalar modes (using the Mukhanov variable $v_S$) reads as \cite{Cailleteau:2011kr}:
\begin{equation}
\frac{d}{d\eta^2}v_S-\Omega \nabla^2v_S- \frac{z_S ^{''}}{z_S}v_S=0,  
\label{scalar}
\end{equation}
where $z_S = a \frac{\dot{\varphi}}{H}$ and 
\begin{equation}
\Omega =1-2\frac{\rho}{\rho_c}.
\label{OmegaDef}
\end{equation}
A prime indicates a derivative with respect to the conformal time $\eta$ while a dot indicates a derivative with respect 
to the  cosmic time $t$. The conformal time is related to the cosmic
time by $\eta = \int \frac{dt}{a(t)}$.
Based on the Mukhanov variable $v_S$, the scalar curvature $\mathcal{R} = \frac{v_S}{z_S}$ can be computed.  
Similarly, for tensor modes \cite{Cailleteau:2012fy}:
\begin{equation}
\frac{d}{d\eta^2}v_T-\Omega \nabla^2v_T- \frac{z_T ^{''}}{z_T}v_T=0, 
\label{tensor}
\end{equation}
where $z_T = a/\sqrt{\Omega}$. The $v_T$ variable relates to amplitude of the tensor modes $h$
through $v_T = \frac{a h}{\sqrt{16\pi G} \sqrt{\Omega}}$. In what follows we will work mainly 
with the $\phi$ variable defined as $\phi :=  h \sqrt{16\pi G}$.

\begin{figure}[ht!]
\centering
\includegraphics[width=7cm,angle=0]{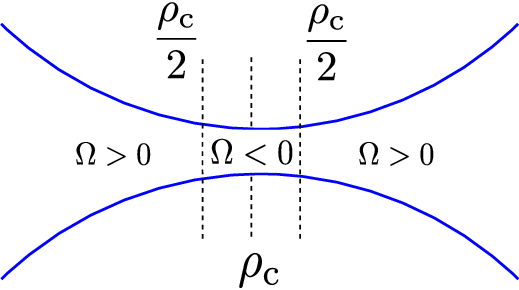}
\caption{Regions of the positive and negative values of the parameter $\Omega$ in a 
typical symmetric cosmic bounce. The silent initial conditions investigated in this paper 
are imposed at $\rho=\rho_c/2$.} 
\label{domains}
\end{figure}

In both cases, there is a $\Omega$ factor in front of the Laplace operator, which is related
with the speed of propagation $c_s$ by $c_s^2 = \Omega$.  When approaching the beginning of 
the Lorentzian  stage the  factor $\Omega$ tends to zero. Because the space dependence is 
suppressed, the different space points are decoupled and become independent one from another. 
This behavior agrees with the predictions of the Belinsky, Khalatnikov and Lifshitz (BKL) conjecture 
\cite{BKL} which states, in particular, that  near a classical  space singularity, different points do 
decouple one from the other.   When $\Omega$ becomes negative one enters the Euclidean 
regime. In the case of a symmetric bounce, the domains of positive and negative values of the 
parameter $\Omega$ are shown in Fig. \ref{domains}.

In this article, for simplicity, we will only study tensor modes. However, because of the 
similarities between Eq. (\ref{scalar}) and Eq. (\ref{tensor}), it is reasonable to expect that 
most of the results will hold also for scalar modes. 

The organization of the paper is as follows. In Sec. \ref{QGP}, general considerations regarding the
generation of quantum tensor perturbations in presence of holonomy corrections are presented. 
The evolution of the horizon as well as the dynamics of the modes are investigated. We also derive 
the equation of motion governing the evolution of the power spectrum. In Sec. \ref{Vac}, a possibility of 
defining a proper vacuum state in both the Lorentzian and  the Euclidean regions is investigated.  
In Sec. \ref{SilenceSec}, the properties of the ``silent surface'', defined as the interface between the Lorentzian and 
Euclidean regions, are analyzed in details. In particular, the possibility of a  ``white noise''-type fluctuations at 
the surface is addressed. The solutions to EOMs in vicinity of the ``silent surface'' are presented as well.
In Sec. \ref{TransPlanckian},  we consider the possibility of 
imposing the initial conditions for $k< m_{\text{Pl}}$ and $k > m_{\text{Pl}}$ separately. Namely, for $k< m_{\text{Pl}}$ 
the initial conditions are imposed at $\rho=\rho_{\text{c}}/2$ while for $k> m_{\text{Pl}}$ the initial conditions are 
imposed at the trans-Planckian surface. We show that a flat power spectrum can be generated from the 
trans-Planckian initial conditions if an appropriate inflationary period takes place. In Sec. \ref{SuperPlanckFlat}, 
we investigate a possible conversion of the $\mathcal{P} \propto k^3$ spectrum generated from  modes with 
$k< m_{\text{Pl}}$ at the initial surface to a scale-invariant shape.  We find that it is indeed possible --but quite difficult-- 
by combining  two evolutions characterized by two different barotropic indices $w_1$ and $w_2$. The resulting 
spectrum is, however, modulated by acoustic oscillations due to the transitional sub-horizon evolution. 
Finally, in Sec. \ref{Summ}, results of the paper are summarized and conclusions are derived. 

\section{Quantum generation of perturbations} \label{QGP}

\subsection{Horizon}

In this paragraph, we study the general behavior of the horizon when the Universe, in its quantum regime 
(effectively described by the holonomy-corrected Hamiltonian), is filled with a barotropic fluid. This is relevant for setting the vacuum for perturbations. \\

The evolution of cosmological perturbations strictly depends on the length of the 
mode under consideration. In particular, there are two regimes (super-Hubble and 
sub-Hubble) in which the evolution of perturbations qualitatively differs. 
The difference is transparent in the Fourier space, where an explicit
wave-number dependence enters the equations of motion.  

Performing the Fourier transform of the perturbation field $v=\int \frac{d^3k}{(2\pi)^3}v_{\bf k} e^{i{\bf k\cdot x}}$,  
the equation of motion satisfied by the Fourier component $v_{\bf k}$ is  
\begin{equation}
\frac{d}{d\eta^2}v_{\bf k}+\left(\Omega k^2- \frac{z ^{''}}{z}\right) v_{\bf k}=0. 
\label{Ftensor}
\end{equation}
Here, and in the rest of this work, for the sake of simplicity, we denote $v:=v_T$ and $z:=z_T$. 

Due to presence of the $\Omega$ factor in Eq. \ref{Ftensor}, the super-Hubble 
and sub-Hubble regimes are {\it not} the same as in the classical case (in the latter case, the 
borderline is approximately given by the Hubble radius $\sim 1/H$). In this new framework, 
the sub-Hubble modes are such that $\left|\Omega k^2\right| \gg \left| \frac{z^{''}}{z} \right|$
while for, the super-Hubble ones, $\left|\Omega k^2\right| \ll \left| \frac{z^{''}}{z} \right|$. The characteristic 
scale of the horizon 
\begin{equation}
\lambda_H = \frac{a}{k_H},
\end{equation}
can be now defined  by the following condition:
\begin{equation}
\left|\Omega k_H^2\right|= \left| \frac{z^{''}}{z} \right|. 
\end{equation}
With this definition, the modes are called super-Hubble if $\lambda \gg \lambda_H$ and  
sub-Hubble if  $\lambda \ll \lambda_H$.  At the sub-Hubble scales the perturbations
are decaying while at the super-Hubble scales, the amplitude of the perturbations is preserved 
(perturbations are ``frozen''). This behavior has important consequences at the quantum level, 
where so-called mode functions (which parametrize the quantum evolution) satisfy the same 
equation as $v_{\bf k}$. \\

Let us now investigate the behavior of $\lambda_{H}$ in the case of a universe filled with 
barotropic matter with a fixed ratio between pressure and density, that is $P=w \rho$, with $w=$ const.
In this case, it can be shown that
\begin{eqnarray}
\frac{z^{''}}{z} &=& \rho_c \kappa a^2 \frac{x}{\Omega^2} \left[ \frac{1}{2}\left(\frac{1}{3}-w\right)
+\left( \frac{7}{6}+9w+\frac{9}{2} w^2 \right)x  \right. \nonumber \\
&+&\left. \left(\frac{7}{3}-9w-6w^2\right)x^2+\left(-\frac{5}{3}+4w+3w^2  \right) x^3 \right],   \nonumber \\ 
\label{zbisz}
\end{eqnarray}
where $x:=\rho/\rho_c$ and $\kappa= 8\pi G$. Because of the $\Omega^2$ factor 
in the denominator, one may expect a divergence when the state of silence 
($\Omega \rightarrow 0$) is reached. For special case of $w=-1$ the Eq. \ref{zbisz} 
reduces to 
\begin{equation}
\frac{z^{''}}{z} = \frac{2}{3} \rho_c \kappa a^2 x(1-x), 
\end{equation} 
which is regular across $\Omega=0$. 

\begin{figure}[ht!]
\centering
\includegraphics[width=7cm,angle=0]{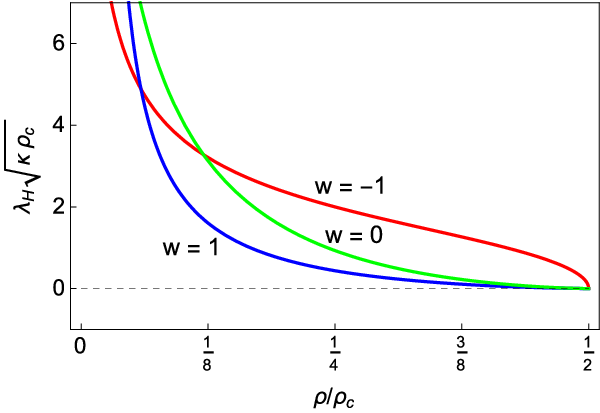}
\caption{Scale of the horizon $\lambda_{H}$ as a function of $\rho/\rho_c$ for different values of the 
barotropic index $w$. Only the Lorentzian domain ($\Omega>0$) is considered. } 
\label{LH}
\end{figure}

The behavior of $\lambda_{H}$ in the Lorentzian domain is plotted in Fig. \ref{LH}.
For $1 \geq  w \geq  -1$, that is for usual types of matter, all wavelengths are becoming 
super-Hubble when $\Omega \rightarrow 0^{+}$  (corresponding to $\rho \rightarrow \frac{\rho_c}{2}^{-}$), 
i.e. $\lambda_H \rightarrow 0$.

\subsection{Quantization of modes}

In this subsection, we analyze the decomposition of perturbations for the mode functions when taking 
into account the specific structure of the holonomy-corrected algebra. This is fundamental for 
the quantum treatment of the problem.\\

The equations of motions (\ref{scalar}) and (\ref{tensor}) are the effective equations including quantum gravity effects. 
Although the quantum effects enter explicitly the equations 
of perturbations (through the factor $\Omega$), only the background degrees of freedom 
were here quantized using the loop approach. The phase space of the perturbed variables remains classical but 
should, in principle, be modified by the quantization. A fully consistent procedure of quantization would require the application of 
the loop quantization to the perturbative degrees of freedom as well. However, for  sufficiently 
large modes ($\lambda \gg l_{\text{Pl}}$), loop quantization should reduce to the canonical one. 

In this study, the Fourier modes $v_{\bf k}(\eta)$ are quantized following the standard 
canonical procedure. Promoting this quantity to be an operator, one performs 
the decomposition
\begin{equation}
\hat{v}_{\bf k}(\eta) = i^{\frac{1}{2}(1-\text{sgn} \Omega)}(f_k(\eta) \hat{a}_{\bf k} +f_k^*(\eta) \hat{a}^{\dagger}_{\bf-k}),
\label{ModeDecomp}
\end{equation}
where $f_k(\eta)$ is the so-called  mode function which satisfies the same equation as 
$v_{\bf k}(\eta)$. Because we are working in the Heisenberg picture, the operator $\hat{v}_{\bf k}(\eta)$
is time dependent, which is encoded in mode functions $f_k(\eta)$.  The creation ($\hat{a}^{\dagger}_{\bf k}$) 
and annihilation ($\hat{a}_{\bf k}$) operators fulfill the commutation relation $[\hat{a}_{\bf k},\hat{a}^{\dagger}_{\bf q}]
=\delta^{(3)}({\bf k-q})$ and are defined at some initial time. 

The new factor $i^{\frac{1}{2}(1-\text{sgn} \Omega)}$ in Eq. (\ref{ModeDecomp}) is due to reality condition
for the field $\phi = v/z$, which leads to $\hat{\phi}_{\bf k}^{\dagger} = \hat{\phi}_{-\bf k}$.  This does not 
translate into a similar condition for $\hat{v}_{\bf k}$ because $z = a/\sqrt{\Omega}$ can be both real and imaginary 
depending on the sign of $\Omega$.

The mode functions are fulfilling the Wronskian condition
\begin{equation}
f_k (f_k')^*-f^*_k f_k^{'} =i, 
\label{wronskian}
\end{equation}
which was proven to keep its classical form \cite{Mielczarek:2013xaa}.\\

As we deal with linear perturbations, which lead to 
Gaussian fluctuations, all the statistical information about the 
structure of the fluctuations is contained in the two-point correlation function. 
For the field $\phi=v/z$, the two-point correlation function is given by:  
\begin{eqnarray}
G\left(r\right)&:=&\langle 0|  \hat{\phi}({\bold x},\eta) \hat{\phi}({\bold y},\eta)| 0 \rangle 
=\int_0^{\infty} \frac{dk}{k} \mathcal{P}_{\phi}(k,\eta) \frac{\sin kr}{kr},
\label{CorrPower} 
\end{eqnarray}
where the power spectrum is
\begin{equation}
\mathcal{P}_{\phi}(k,\eta) = \frac{k^3}{2 \pi^2}  \left| \frac{f_k}{z}\right|^2, 
\label{spectrum}
\end{equation}
and $r= |{\bold x}-{\bold y}|$.

\subsection{Equation of motion for the power spectra} 

In this subsection, we study how the spectrum, as defined in the previous section, can be propagated in time.\\

The idea is to  derive the equation governing the evolution of the power spectrum defined by Eq. (\ref{spectrum}).
Usually, the evolutions of $v_k$  and $z$ are  calculated first and the power spectrum is derived subsequently. 
However, in some cases, it is possible and useful to obtain directly the evolution equation of  $\mathcal{P}(k)$. 
In particular, this is relevant when initial conditions are determined by the form of the correlation function $G(r)$. 
This situation appears when imposing initial conditions at the ``silent surface".

By using the equation of motion for $v_k$, as well as  the Wronskian condition (\ref{wronskian}), 
one can show that the  power spectrum fulfills the following nonlinear differential equation:
\begin{equation}
\frac{d^2 \mathcal{P}}{d\eta^2}-\frac{1}{2\mathcal{P}} \left( \frac{d \mathcal{P}}{d\eta} \right)^2
+2\Omega k^2 \mathcal{P}+2 \frac{d\mathcal{P}}{d\eta} \frac{z'}{z}
-\frac{1}{2z^4 \mathcal{P}} \left( \frac{k^3}{2\pi^2} \right)^2=0.   
\label{Ermakow}
\end{equation}
After a change of variables, this equation can be reduced to the so-called Ermakov equation.

Equation (\ref{Ermakow}) can be written as a set of two first order differential equations. 
The advantage of this decomposition is that the obtained system of equations is free from
the divergence at $\Omega=0$. This leads to:
\begin{eqnarray}
\frac{d\mathcal{P}}{d\eta}  &=&  \frac{\mathcal{G}}{a^2} \Omega, \label{EqP} \\
\frac{d\mathcal{G}}{d\eta} &=& -2(ak)^2 \mathcal{P}
+\frac{\Omega}{2a^2\mathcal{P}}\left[\mathcal{G}^2+\left( \frac{k^3}{2\pi^2} \right)^2\right]. 
 \label{EqG}
\end{eqnarray}
Importantly, the fixed point ($\mathcal{P}'=0,\mathcal{G}'=0$) of this set of equations is given by 
\begin{eqnarray}
\mathcal{G} &=& 0,  \\
\mathcal{P} &=&  \left(  \frac{k}{2\pi}\right)^2 \frac{\Omega}{a^2}, 
\end{eqnarray}
which agrees with the $\Omega$-corrected Minkowski vacuum (\ref{OmegaMinkowski}).

\section{Vacuum} \label{Vac}

In this section, we make some important remarks on how one can define a vacuum state in 
both Lorentzian and Euclidean sector.\\

A first possibility to evaluate the spectrum in holonomy-corrected effective 
loop quantum cosmology ($\Omega$-LQC) is to set initial conditions in the 
remote Lorentzian past (of the contracting branch) and calculate the resulting 
spectrum. Such case has been investigated before \cite{barrau_spectrum}. 
This is mathematically tantalizing and probably consistent but propagating perturbations 
through the Euclidean phase where there is, strictly speaking, no time anymore, 
is questionable. The main aim of this article is to investigate the possibility of 
imposing initial conditions at the interface between the Lorentzian and Euclidean 
regions. \emph{A priori} nothing is know about the state of perturbations at this 
moment in time. This is the same difficulty as for initial conditions 
for the inflationary perturbations in the standard approach. 

The usual assumption at this point is that the perturbations 
are initially in their vacuum state. Such an ansatz can be of course questioned. 
It is, however, a reasonable choice and it is worth investigating 
its consequences. In particular, within the standard inflationary evolution, 
the assumption of an initial vacuum state leads to a power spectrum 
being in agreement with cosmological observations.

In principle, one could consider the perturbation fields classically and 
ignore quantization issues. However, in that case, no normalization of 
the modes would be available. Beyond its legitimacy, taking into account 
the quantum evolution of perturbations is, therefore, heuristically important.\\

We investigate the vacuum state for the holonomy-corrected 
case. To do so, the Hamiltonian for the considered type of perturbations has 
to be clearly defined. We adopt here observations made in \cite{Mielczarek:2013xaa}, 
where it has been shown that the equation of motion (\ref{tensor}) can be 
recovered by considering a wave equation on the effective metric 
\begin{equation}
g^{eff}_{\mu\nu}dx^{\mu} dx^{\nu} =  -\sqrt{\Omega} a^2 d\eta^2
+\frac{a^2}{\sqrt{\Omega}} \delta_{ab}da^adx^b.  \label{effectivmetric}
\end{equation}
However, it is worth mentioning that this effective metric is only an 
auxiliary object and may not have physical relevance due to 
the fact that the constraints algebra is subject of deformations, 
as it will be discussed in Sec. \ref{TransPlanckian}.  Therefore,   
the divergence of auxiliary effective metric at $\Omega=0$ seems 
to not to be a problematic issue. 

Based on Eq. \ref{effectivmetric} the action for the massless field $\phi = v/z$ is given by
\begin{eqnarray}
S &=& - \frac{1}{2} \int d^4 x \sqrt{-g} g^{\mu\nu} \partial_{\mu}\phi \partial_{\nu}\phi= \int d\eta L  \nonumber \\
&=&  \frac{1}{2} \int d\eta d^3x a^2\left[ \frac{1}{\Omega} \left(  \phi' \right)^2-\left( \partial_i \phi \right)^2\right] \nonumber \\   
&=&   \frac{1}{2} \int d\eta d^3x \left[ v'^2- \Omega \left( \partial_i v \right)^2+v^2  \frac{z^{''}}{z}\right] ,  
\end{eqnarray}   
where an integration by parts was used in the second equality. Using the canonical 
momenta $\pi = \frac{\delta L}{\delta v'} = v'$, the Hamiltonian can be defined
\begin{eqnarray}
H = \int d^3x \pi v' -L  
&=\frac{1}{2} \int d^3x\left[ \pi^2+\Omega\left( \partial_i v \right)^2-v^2  \frac{z^{''}}{z}\right].  
\end{eqnarray} 
With mentioning that this Hamiltonian can be also obtained directly from the original 
LQC Hamiltonian for perturbations, without referring to the effective metric (\ref{effectivmetric}). 

The quantum version of this Hamiltonian can be written as  
\begin{eqnarray}
\hat{H} = \frac{1}{2}\int d^3k\left[ \hat{a}_{\bf k}\hat{a}_{-\bf k} F_{k} + 
\hat{a}^{\dagger}_{\bf k}\hat{a}^{\dagger}_{-\bf k}F^*_k 
+ \left(2 \hat{a}^{\dagger}_{\bf k}\hat{a}_{\bf k}+ \delta^{(3)}(0)\right)E_k\right], 
\end{eqnarray}
where 
\begin{eqnarray}
F_k &=& (f'_k)^2+\omega_k^2f_k^2, \\
E_k &=& |f_k'|^2+\omega_k^2|f_k|^2, 
\end{eqnarray}
and 
\begin{equation}
\omega_k^2 = \Omega k^2 -\frac{z^{''}}{z}. 
\label{omega2}
\end{equation} 
The vacuum expectation value is
\begin{eqnarray}
\langle 0 | \hat{H} | 0 \rangle = \delta^{(3)}(0) \frac{1}{2}\int d^3k E_k. 
\end{eqnarray}
Except the case of compact spatial volume the value of $\langle 0 | \hat{H} | 0 \rangle $ is 
expected to be infinite. However, it cannot be just simply subtracted as in case of the  
Minkowski space-time because $E_k$ are now time dependent functions. Furthermore, 
finiteness of the integral is expected due to a cut-off at the Planck scale energies. 

The ground state (vacuum) can be found by minimizing $E_k$ while taking the
Wronskian condition (\ref{wronskian}) into account. This leads to the condition that 
the energy can be minimized if and only if $\omega_k^2>0$.  Then, the
interpretation of the excitations of the field as particles is also possible.
In that case, the corresponding  vacuum state is 
\begin{equation}
f_k = \frac{e^{-i \omega_k \eta}}{\sqrt{2\omega_k}}. 
\label{VacuumMode}
\end{equation}
This is rigorously satisfied only if $\omega_k=$ const, which is not always the case. 
However, if $\omega_k$ is a slowly varying function of time,  Eq. (\ref{VacuumMode})
remains a good approximation of the vacuum state.\\ 

The positivity of $\omega^2_k$, required for a proper definition of the vacuum state,  depends 
on both the sign of $\Omega$ and on the value of $\frac{z^{''}}{z}$. In the Lorentzian regime ($\Omega>0$) it is always 
possible to find values of $k$ for which $\omega^2_k$ is positive. If $\frac{z^{''}}{z}$ is negative, this 
is the case for any $k$. On the other hand, if $\frac{z^{''}}{z}$ is positive, this requires sufficiently 
large (sub-Hubble) $k$-valued mode. 

The situation, however, changes in the Euclidean regime where $\Omega<0$. Now, the $k^2$ term
in the $\omega^2_k$ function is multiplied by a negative factor, and the positivity of $\omega^2_k$
can be satisfied only if $\frac{z^{''}}{z}$ takes a negative value.\\

\begin{figure}[ht!]
\centering
\includegraphics[width=7cm,angle=0]{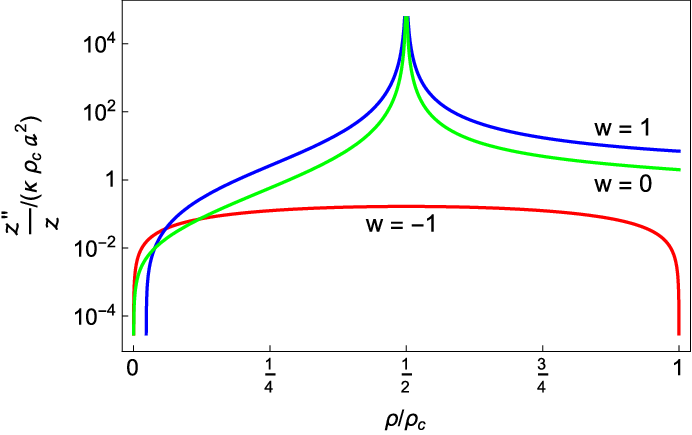}
\caption{The $\frac{z^{''}}{z}$ as a function of $\rho/\rho_c$ for different values of $w$.} 
\label{zbarz}
\end{figure}

In Fig. \ref{zbarz}, we show $\frac{z^{''}}{z}$ as a function of the energy density for some representative
values of the barotropic index $w$.  
One can show that for $1 \geq  w \geq  -1$, the $\frac{z^{''}}{z}$ term is positive in the 
whole Euclidean domain. There is, therefore, no well defined vacuum state in this case. 

\begin{figure}[ht!]
\centering
\includegraphics[width=7cm,angle=0]{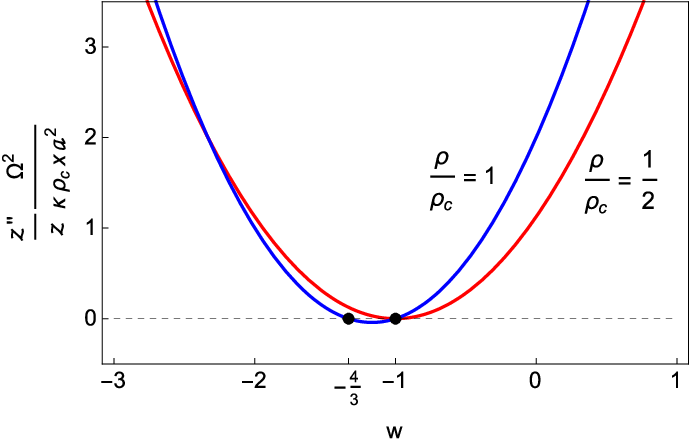}
\caption{Appropriately rescaled $\frac{z^{''}}{z}$ as a function of the parameter $w$ at 
the boundaries of the Euclidean domain ($\frac{\rho}{\rho_c}=1/2$ and $\frac{\rho}{\rho_c}=1$). 
The dots indicate boundary regions of the negative $\frac{z^{''}}{z}$  for  $\frac{\rho}{\rho_c}=1$.} 
\label{VaccumCC}
\end{figure}

\begin{figure}[ht!]
\centering
\includegraphics[width=6cm,angle=0]{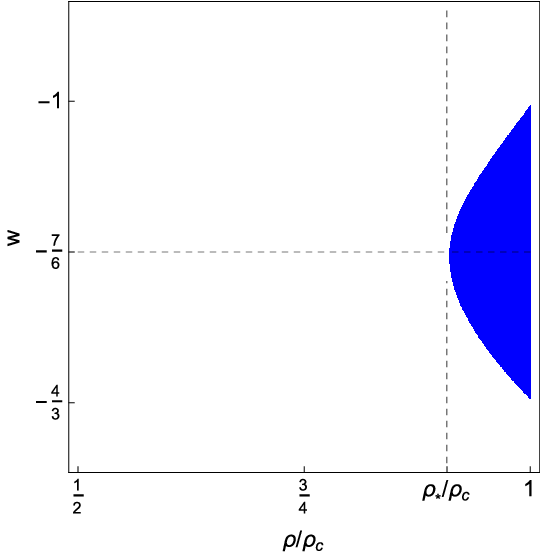}
\caption{The shadowed (blue) area represents region of negative values of $\frac{z^{''}}{z}$ in the Euclidean sector.} 
\label{RegVac}
\end{figure}

In order to investigate whether there are possible domains of negative $\frac{z^{''}}{z}$  beyond 
the  $1 \geq  w \geq  -1$ region let us firstly study the value of the $\frac{z^{''}}{z}$  as a function 
$w$ at limiting energy scales of the Euclidean region (see Fig.~\ref{VaccumCC}).
By analyzing Fig.~\ref{VaccumCC} one can conclude that  $\frac{z^{''}}{z}$ is positive definite 
at the signature change surface, where $\Omega=0$ and $\frac{\rho}{\rho_c}=1/2$. On the 
other hand, at the highest energy density point ($\frac{\rho}{\rho_c}=1$) the $\frac{z^{''}}{z}$ 
function is becoming negative for $-1>w>-\frac{4}{3}$, which corresponds to the so-called 
phantom sector. In that case, the vacuum can be defined at super-Hubble scales. Once can 
now ask if this regime extends to the lower energy density scales. By analysing the formula (\ref{zbisz})
one can find that the region of negative values of $\frac{z^{''}}{z}$ extends to the energy 
density $\rho_* \approx 0.90784 \rho_c$ and is centered around $w=-\frac{7}{6}$. The  
region at the $\left(\frac{\rho}{\rho_c},w\right)$ plane is presented in Fig. \ref{RegVac}. 

The fact that the vacuum state can be well defined at  super-Hubble scales is a new feature.  
However, one has to keep it mind that it is possible to achieve for certain phantom-like 
values of $w$ and only in vicinity of the highest energy density region. 

One can speculate that universe can ``tunnel'' from such a vacuum state (defined at the large scales
in the interior of the Euclidean sector) to the Lorentzian sector, defining structure of the super-Hubble 
inhomogeneities at the onset of the Lorentzian sector. The process of tunneling which we have in 
mind here is related with the fact that the mode equation (\ref{Ftensor}) can be perceived as a 
Schr\"odinger equation with $f_k$ playing a role of the wave function and where the conformal ``time'' $\eta$ 
plays a role of spatial direction. Furthermore the ``particle" energy minus potential energy is then given by 
\begin{equation}
E-V = \omega^2_k =  \Omega k^2 -\frac{z^{''}}{z}.
\end{equation}
Without the lose of generality, we can talk about the zero energy ``particle" for which the potential 
is just  $V = -\omega^2_k =-\Omega k^2 +\frac{z^{''}}{z}$. Furthermore, the analogy is justified by the
fact that in the Euclidean sector the conformal ``time'' $\eta$ does not parametrize evolution but is 
treated at the similar footing as the remaining spatial coordinates (at least while linear perturbations
are considered). Then the tunneling process considered here is very similar to the one studied by 
A.~Vilenkin (see Ref. \cite{Vilenkin:1983xq}). Here, similarly to the quantum cosmological case (described 
by the WDW equation) there is no external time in which the tunneling process occurs. What actually 
happens is that the regularity of the wave function implies its spreading across the potential barrier, such 
that with some non-vanishing probability the system can be localized in the Lorentzian sector. But 
when already beyond the Euclidean domain, the meaning of $\eta$ changes and cosmological evolution  
start to move away the system from the Euclidean domain. Further detailed analysis is however needed 
to approve viability of the described process.  

To conclude, under assumption of the scenario described above, the super-Hubble power 
spectrum in vicinity of the silent surface has chance to scale as $k^3$, as we partially demonstrate below. 
 
\subsection{Vacuum at $\Omega=-1$}

As an example, let us consider the state of vacuum in the case $w=-\frac{7}{6}$ and  $\Omega=-1$, 
for which 
\begin{equation}
\frac{z^{''}}{z} = -\frac{1}{24} \rho_c \kappa a^2.
\end{equation}
Based on this,  the factor (\ref{omega2}) can be written as 
\begin{equation}
\omega_k^2=-k^2+\frac{1}{24} \rho_c \kappa a^2,  
\end{equation}
which is positive for wavelength $\lambda > \lambda_H = \frac{2\sqrt{6}}{\sqrt{\kappa \rho_c}}$.
The corresponding vacuum normalization of the mode function  is then given by 
\begin{equation}
|f_k|^2 =\frac{1}{2 \omega_k}  = \frac{\sqrt{6}}{\sqrt{\kappa \rho_c } a}. 
\end{equation}
Subsequently, the power spectrum can be written as follows
\begin{equation}
\mathcal{P}_{\phi}(k) \equiv \frac{k^3}{2\pi^2}\left| \frac{f_k}{z}\right|^2 =
\frac{\sqrt{6}}{2 \pi^2 \sqrt{ \kappa \rho_c }}\left( \frac{k}{a}\right)^3 \propto k^3.      
\end{equation}
This justifies (partly) the assertion presented at the end of the previous subsection. 
The power spectrum differs from the standard Bunch-Davies case, for which $\mathcal{P}_{\phi} \propto k^2$.
Using the definition (\ref{CorrPower}), one can show that the corresponding  correlation function is vanishing 
$\langle 0|  \hat{\phi}({\bold x},\eta) \hat{\phi}({\bold y},\eta)| 0 \rangle=0$. In this case, fluctuations have a 
white noise spectrum. This is intuitively compatible with the idea that space points are decorrelated. 

\subsection{Vacuum at $\Omega \approx 0$} 

This case is the most relevant one for the subject of this article. However, except of the very 
specific equation of state with $w=-1$,  $\omega^2_{k}$ is divergent when $\Omega \rightarrow 0$. 
Furthermore, as we have shown, there are no values of $w$ for which the value of  $\frac{z^{''}}{z}$ 
is negative at $\Omega=0$ or in vicinity of this point. Therefore, for the moment it seems 
that there is no direct way to associate vacuum state with the signature change. The \emph{state of 
silence} at  $\Omega = 0$ which we are going to introduce later is, therefore, most probably not 
a vacuum state (or not a vacuum state in the sense considered here). 
 
\subsection{Vacuum at $\Omega>0$} 

The $\Omega>0$ region is where initial conditions for the quantum fluctuations are 
usually imposed. In this case, the vacuum sate is well defined for $\lambda \ll \lambda_H$,
where $\omega^2_k \approx \Omega k^2$. This, applied to Eq. (\ref{VacuumMode}), 
leads to the following expression for the vacuum normalization:
\begin{equation}
|f_k|^2 = \frac{1}{2k \sqrt{\Omega}}.
\end{equation}
This normalization has also been derived using independent arguments \cite{Mielczarek:2013xaa,Barrau:2014maa}.
The corresponding power spectrum is 
\begin{equation}
\mathcal{P}_{\phi}(k) \equiv \frac{k^3}{2\pi^2}\left| \frac{f_k}{z}\right|^2 =  
\left(\frac{k}{2\pi}\right)^2  \frac{\Omega}{a^2} \propto k^2.   
\label{OmegaMinkowski}
\end{equation}
This is the $\Omega-$deformed Bunch-Davies vacuum. In that case the holonomy corrections change the normalization but do not modify the shape of the standard spectrum.

\section{Physics at the surface of initial conditions} \label{SilenceSec} 

In this section, we derive different relations useful for calculating the spectra. 
In particular, initial values for $\mathcal{P}$ and $\mathcal{G}$ governed by 
Eqs. (\ref{EqP}) and (\ref{EqG}) will be studied.

\subsection{Solution for $\Omega \approx 0$}

In the vicinity of the interface between the Euclidean and the Lorentzian regions (where $\Omega=0$), 
the equation of motion for the $\phi$ variable, 
\begin{equation}
\phi^{''}+\left(2\mathcal{H}- \frac{\Omega'}{\Omega}\right)\phi'-\Omega \Delta \phi = 0,
\label{EqPhiOmega}  
\end{equation}
simplifies to 
\begin{equation}
\phi^{''}+\left(2\mathcal{H}- \frac{\Omega'}{\Omega}\right)\phi' \approx 0. 
\label{phiapprox} 
\end{equation}   
Despite this approximation, the equation remains singular due to presence of 
the $\frac{\Omega'}{\Omega}$ factor. The solution across the  silent surface 
is however regular. In order to find the solution, the equation 
(\ref{phiapprox}) can be integrated to 
\begin{equation}
\phi' = c_1 \frac{\Omega}{a^2}.  
\end{equation}
Further integration leads to the solution 
\begin{equation}
\phi=c_2+c_1 \int^{\eta} \frac{\Omega}{a^2}d\eta' = c_2+c_1 \int^{\eta} \frac{d\eta'}{z^2}.  
\end{equation}
Because the $\Omega$ factor appears only in the numerator, no pathological behavior is to be expected. 
Using the above analysis, the solution to the  simplified equation for the mode functions   
\begin{equation}
\frac{d}{d\eta^2}f_k-\frac{z^{''}}{z}f_k=0, 
\end{equation}
takes the following form:
\begin{equation}
f_{k} = z\left(A_k +B_k  \int_{\eta_0}^{\eta} \frac{d\eta'}{z^2} \right).
\label{fkOmega=0sol}
\end{equation}
It is worth stressing that evolution of the amplitude $f_k/z$ is regular through the 
silent surface (when $\Omega=0$) despite the fact that the $\frac{z^{''}}{z}$ factor 
present in the equation for $f_k$ is generically divergent. The origin of this divergence is the $\sqrt{\Omega}$ factor 
occurring in definition of $z$. The  $\sqrt{\Omega}$ factor is non-differentiable at $\Omega=0$ leading 
to divergences occurring in equations governing the evolution of the mode functions. However, the $\sqrt{\Omega}$ 
factor does not appear in the equations for amplitudes (such as Eq. (\ref{EqPhiOmega})), that have 
regular solutions.

The $A_k$ and $B_k$ in Eq. (\ref{fkOmega=0sol}) are constants of integration and fulfill the following relation:
\begin{equation}
A_kB^*_k-A^*_kB_k=i,
\label{WronskianAB}
\end{equation}
due to the Wronskian condition. This leads to
\begin{eqnarray}
 \left|\frac{f_k}{z}\right|^2 &=& |A_k|^2+|B_k|^2 \left( \int_{\eta_0}^{\eta} \frac{d\eta'}{z^2}\right)^2 \nonumber \\
 &+&(A_kB^*_k+A^*_kB_k) \int_{\eta_0}^{\eta} \frac{d\eta'}{z^2}.
\label{initialamplitude}
\end{eqnarray}

\subsection{Initial conditions for the perturbations} 

One can now use Eq. (\ref{initialamplitude}) with $\eta_0=0$ corresponding to the 
transition point $\Omega=0$  in order to impose initial conditions. The initial 
conditions for the fields $\mathcal{P}$ and $\mathcal{G}$ can then be written as follows:
\begin{eqnarray}
\left. \mathcal{P} \right|_{\eta=0} &=& \frac{k^3}{2\pi^2} |A_k|^2,    \\
\left. \mathcal{G} \right|_{\eta=0} &=&  \frac{k^3}{2\pi^2} \left( A_kB^*_k+A^*_kB_k\right).
\end{eqnarray}

By expressing $A_k$ and $B_k$ in terms of amplitudes and phases 
\begin{eqnarray}
A_k &=& \tilde{A}_k e^{i\alpha}, \\
B_k &=& \tilde{B}_k e^{i\beta},
\end{eqnarray} 
the Wronskian condition (\ref{WronskianAB}) leads to 
\begin{equation}
2  \tilde{A}_k\tilde{B}_k \sin(\alpha-\beta)=1,
\end{equation}
and 
\begin{equation}
A_kB^*_k+A^*_kB_k = 2 \tilde{A}_k\tilde{B}_k \cos(\alpha-\beta)= \cot(\alpha-\beta).
\end{equation} 
The cotangent function has a period of $\pi$. Let us define the phase 
difference $X = \alpha - \beta \in (0,\pi)$. For the particular value  $X= \pi/2$ we have 
\begin{equation}
A_kB^*_k+A^*_kB_k = 0,
\end{equation}
so  $\left. \mathcal{G} \right|_{\eta=0}$ is vanishing. 

If we assume that the phase difference has a flat distribution  
\begin{equation}
P(X) = \frac{1}{\pi},
\end{equation}
then the distribution of the values of 
\begin{equation}
Y = \cot (X)
\end{equation}
is given by a Cauchy distribution
\begin{equation}
P(Y) = \frac{1}{\pi} \frac{1}{(1+Y^2)},
\end{equation}
which is peaked at $Y=0$. This might be seen as a probabilistic 
motivation for choosing $\left. \mathcal{G} \right|_{\eta=0}=0$. In other words, 
if the phase difference $X$ is chosen randomly, then the most probable 
value of  $\left. \mathcal{G} \right|_{\eta=0}$ is  zero. This is not a demonstration but rather an heuristic argument for this choice.
We will use this value to perform numerical computations in the following. 

\subsection{Correlation functions}
\label{corrFun}

Here, we explicitly show that white noise initial conditions 
at the silent surface lead to a power spectrum cubic in $k$.\\

\begin{figure}[ht!]
\centering
\includegraphics[width=7cm,angle=0]{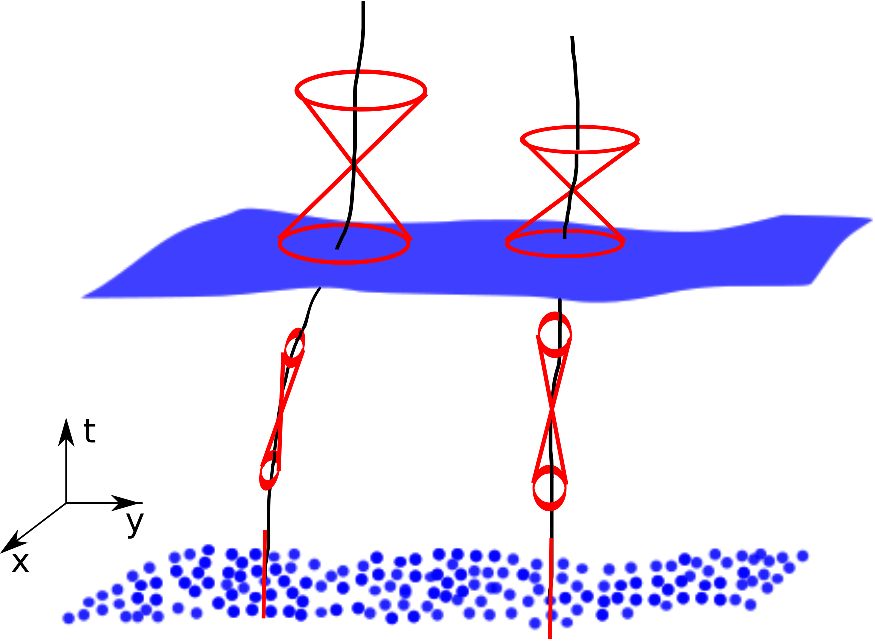}
\caption{Pictorial representation of the evolution of the light-cones towards the 
silence surface $\rho=\rho_c/2$, where space-time becomes a congruence of  
time lines. } 
\label{Silence}
\end{figure}

The state of silence is characterized by the suppression of spatial derivatives, 
which leads to the decoupling of the evolution at different  space points. As already explained before, 
and as shown in \cite{Mielczarek:2012tn}, this 
phase takes place in the vicinity of $\rho=\rho_{\text{c}}/2$.
While approaching the state of silence, light cones collapse onto time lines, 
as pictorially presented in Fig. \ref{Silence}. 
Because the evolutions at different space points are decoupled and run independently, 
one can expect that the correlations between physical quantities evaluated at different points 
are vanishing. More precisely, it is reasonable to expect that there are still some correlations but 
on very small scales, with a correlation length $\xi \sim l_{\text{Pl}}$. This situation can 
be modeled by the following correlation function   
\begin{equation}
G(r) = \left\{ \begin{array}{ccc}    G_0 & \text{for}  & \xi \geq r \geq 0,  \\ 
0 & \text{for} & r>\xi.  \end{array} \right.
\label{corr2}
\end{equation}
Given  a correlation function $G(r)$, the power spectrum can be straightforwardly found by using the relation
\begin{equation}
\mathcal{P}(k) = \frac{2}{\pi} k^3 \int_{0}^{\infty} dr~G(r)~r^2 \frac{\sin(kr)}{kr}. 
\label{PowerCorr}
\end{equation}
For the considered model, using Eq. (\ref{PowerCorr}), we find:
\begin{eqnarray}
\mathcal{P}(k) &=&  \frac{2}{\pi} k^3 G_0 \int_{0}^{\xi} dr~r^2 \frac{\sin(kr)}{kr}  \nonumber \\
&=& \frac{2}{\pi} G_0 \int_{0}^{\xi k} dx~x~\sin x  \nonumber \\
&=& \frac{2 G_0}{\pi } \left[-k\xi \cos k\xi+\sin k\xi   \right].
\end{eqnarray}
In the limit $k\xi \ll 1$:
\begin{equation}
\mathcal{P}(k) =  \frac{2}{3} \frac{G_0}{\pi}  (k\xi)^3 +\mathcal{O}((k\xi)^5). 
\end{equation}
This shows that, at large scales, the power spectrum is of the $k^3$ form, as expected for white noise.

Importantly, $G\left(r\right)=0$ in Eq. (\ref{PowerCorr}) corresponds to the trivial $\mathcal{P}(k)=0$ case. 
However, from the definition Eq. (\ref{CorrPower}) we know that $\mathcal{P}(k)\propto k^3$ also give $G\left(r\right)=0$.

\section{Trans-Planckian modes} \label{TransPlanckian}

The equations governing the evolution of both tensor and scalar perturbations are valid only 
for modes that are larger than the Planck scale. This is because at  short scales the notion
of continuity is expected to break down due to quantum gravity effects. However, some knowledge 
about how the so-called \emph{trans-Planckian} modes behave can be gained by considering 
quantum deformations of space-time symmetries.  

The relevant type of deformations can be inferred from the form of the algebra of quantum-corrected
constraints. For the holonomy corrections considered in this article, the algebra of constraints 
takes the following form \cite{Cailleteau:2011kr}:
\begin{eqnarray}
\left\{D[M^a],D [N^a]\right\} &=& D[M^b\partial_b N^a-N^b\partial_b M^a], \nonumber \\
\left\{D[M^a],S^Q[N]\right\} &=& S^Q[M^a\partial_a N-N\partial_a M^a],  \nonumber  \\
\left\{S^Q[M],S^Q[N]\right\} &=& \Omega D\left[q^{ab}(M\partial_bN-N\partial_bM)\right].  \nonumber
\end{eqnarray} 
The $D$ term is the diffeomorphism constraint and $S^Q$ is the holonomy-corrected scalar 
constraint. The constraints play the role of generators of the symmetries and are parametrized by
the lapse function ($N$) and the shift vector ($N^a$). The $\Omega$ is a deformation factor (given by Eq. (\ref{OmegaDef})), equal one in the classical limit while $q^{ab}$ is the inverse of the spatial metric.

The algebra of constraints reduces to the Poincar\'e algebra, describing isometries of 
the Minkowski space, in the short scale limit. However, due to the deformation of the algebra
of constraints, the Poincar\'e algebra is deformed as well \cite{Bojowald:2012ux, Mielczarek:2013rva}.    
The deformation of the Poincar\'e algebra manifests itself through modifications of the dispersion 
relations. As discussed in \cite{Mielczarek:2013rva}, this leads to an energy 
dependent speed of propagation, such that the group velocity tends to zero when the
energy of the modes approaches $m_{\text{Pl}}$.    

This effect can be introduced by considering a $k$-dependence in the $\Omega$ function. 
For phenomenological purposes, one can assume that the $\Omega$ factor in front of the 
$k^2$ term in the equations of motion  (\ref{Ftensor}) is replaced by
\begin{equation}
\Omega = \left(1- 2\frac{\rho}{\rho_{\text{c}}}\right)\left(1 - \frac{1}{m_{\text{Pl}}^2}  \left( \frac{k}{a} \right)^2\right), 
\label{BetaGeneralized}
\end{equation}  
which is defined for $k \leq a\ m_{\text{Pl}}$.  For $k > a\ m_{\text{Pl}}$ the continuous 
space-time approximation is expected to break down and the equation of motion  (\ref{Ftensor}) cannot be applied.
Another way of introducing this effect is by performing the replacement $(k/a)^2 \rightarrow \Upsilon^2(k)$,
as usually done when studying  modified dispersion relations for the  propagation of cosmological 
perturbations (See \emph{e.g.} \cite{KowalskiGlikman:2000dz}). The $\Upsilon^2(k)$ function
encodes deformation of the dispersion relation.

Eq. (\ref{BetaGeneralized}) can be studied in  different regimes. When considering
large scales ($\frac{k}{a} \ll m_{\text{Pl}}$) the new factor in equation (\ref{BetaGeneralized}) can 
be neglected and $\Omega \approx 1- 2\frac{\rho}{\rho_{\text{c}}} $. This is, in particular, valid for the 
modes characterized by $k\ll m_{\text{Pl}}$ at the surface of initial conditions. On the other hand, when 
we are far away from the initial surface ($\rho \ll \rho_{\text{c}}$), the quantum effects are relevant 
only for short scale modes, then $\Omega \approx 1- \frac{1}{m_{\text{Pl}}^2}\left(\frac{k}{a} \right)^2$.  
Of course when $\rho \sim \rho_{\text{c}}$ 
and $\frac{k}{a} \sim m_{\text{Pl}}$ both effects should be taken into account simultaneously.  
In the classical domain, when $\rho \ll \rho_{\text{c}}$ and 
$\frac{k}{a} \ll m_{\text{Pl}}$, one naturally recovers $\Omega \approx 1$. \\

Initial condition for the modes can be imposed when 
\begin{equation}
k\approx a\ m_{\text{Pl}}. 
\end{equation}
At this time $\Omega \approx 0$ for $\rho \ll \rho_{\text{c}}$ (perhaps also for $\rho \sim \rho_{\text{c}}$
as the modification of the dispersion relation is here taking into account both effects basically independently).
In this limit, the equation of motion  (\ref{Ftensor}) simplifies to 
\begin{equation}
\frac{d}{d\eta^2}v_k-\frac{z^{''}}{z}v_k=0,
\label{ModesTransPlanckianEQ}
\end{equation}  
as in the vicinity of $\rho = \rho_{\text{c}}/2$.  One can therefore expect that the state of asymptotic silence 
is realized also at the scales of the order of the Planck length. \\

\begin{figure}[ht!]
\centering
\includegraphics[width=8cm,angle=0]{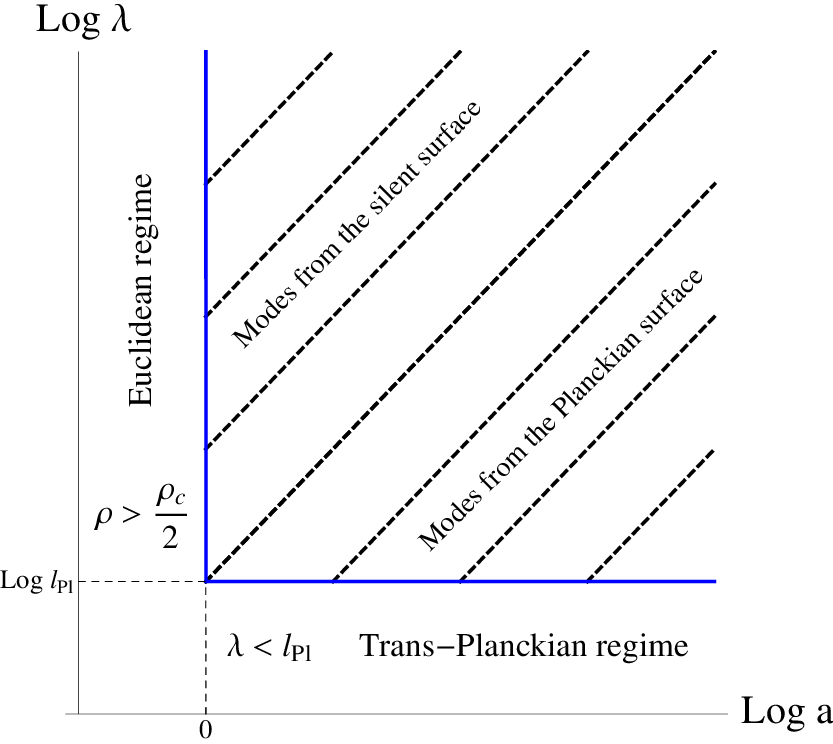}
\caption{The thick (blue) line represents the surface at which the initial conditions are imposed. 
For the modes with $k<m_{\text{Pl}}$ at $a=1$ this is the silent surface while for $k>m_{\text{Pl}}$ 
it is the Planckian surface. The dashed lines represent the evolution of the different length  
scales $\lambda = \frac{a}{k}$. The wavelength $\lambda = l_{\text{Pl}} a$ sets the limit between the 
different types of initial conditions.} 
\label{Boundaries}
\end{figure}

Let us now compute the total power spectrum, including $k< m_{\text{Pl}}$ and $k > m_{\text{Pl}}$ regions. 
Initial conditions for the $k< m_{\text{Pl}}$ modes will be imposed at the silence surface. 
In turn, initial conditions for the $k > m_{\text{Pl}}$ are imposed at the Planckian surface, when 
$k\approx a\ m_{\text{Pl}}.$ In Fig. \ref{Boundaries} the initial value surfaces are presented on the 
$\log \lambda -\log a$ plane.

A mode with a given $k$ exits the Planckian surface when $k \approx a(t_e) m_{\text{Pl}}$,
(with $a(t_0)=1$ at the beginning of the Lorentzian phase).   
For the modes being of the order of the Planck length ($k\approx a\ m_{\text{Pl}}$), 
the equation of motion reduces to the $k$-independent form (\ref{ModesTransPlanckianEQ}). 
Because at $k\approx a\ m_{\text{Pl}}$ and for $\rho \ll \rho_c$ one has
\begin{equation}
\omega_k^2 =-\frac{z^{''}}{z} \approx \rho \kappa a^2 \frac{1}{2}\left(-\frac{1}{3}+w\right),
\end{equation}
the vacuum normalization (\ref{VacuumMode}) can be applied only if $w>\frac{1}{3}$. In this 
case, the spectrum at the Planckian surface is
\begin{equation}
\mathcal{P}(k=m_{\text{Pl}} a) \approx 
\frac{m^3_{\text{Pl}} a^{\frac{3}{2}(1+w)}}{(2\pi)^2 \sqrt{4\pi G \rho_0 |w-1/3|}},
\label{PSilTrans}
\end{equation} 
where $\rho_0$ is matter energy density at $a=1$. In particular, for $w=1$ this leads to
$\mathcal{P}(k=m_{\text{Pl}} a) \propto k^3$. This initial power spectrum cannot be, 
however, converted into a scale-invariant one in the same period, driven by the 
barotropic fluid with $w=1$. A more relevant case for cosmology is the one with
$\mathcal{P}(k=m_{\text{Pl}} a) =$ const. This can be obtained from (\ref{PSilTrans}) 
for $w=-1$. However, in that case, $\omega^2_k<0$ and the spectrum  (\ref{PSilTrans}) 
does not correspond to the initial vacuum state. Nevertheless, this initial state leads 
to predictions being in qualitative agreement with the cosmological observations. 
The state itself is, however, not distinguished at the purely theoretical ground.

The requirement $\mathcal{P}(k = m_{\text{Pl}}a)=$ const at the Planckian surface in fact 
agrees with the standard vacuum-type normalization of modes. To see this explicitly, let us notice
that the evolution of the amplitudes of perturbations can be approximated classically by $\phi_k = \frac{c_k}{a}$ 
(in the regime between the horizon and the Planckian surface). The power spectrum 
$\mathcal{P}(k) = \frac{k^3}{2\pi^2} \frac{|c_k|^2}{a^2}$ at the Planckian surface is therefore 
equal to
\begin{equation}
\mathcal{P}(k=m_{\text{Pl}} a) = \frac{k\ m^2_{\text{Pl}}}{2\pi^2} |c_k|^2. 
\end{equation}
It can be made constant by setting $c_k \sim \frac{1}{\sqrt{k}}$, which 
corresponds to the Bunch-Davies  normalization of modes.\\ 

In order to illustrate the procedure, we have performed numerical computations of the power 
spectra with initial conditions: 
\begin{eqnarray}
\left. \mathcal{P} \right|_{\Omega=0} &=& (k/m_{\text{Pl}})^3,  \\
\left. \mathcal{G} \right|_{\Omega=0} &=& 0,
\end{eqnarray}
for $k< m_{\text{Pl}}$ and 
\begin{eqnarray}
\left. \mathcal{P} \right|_{k = m_{\text{Pl}}a} &=& 1,  \\
\left. \mathcal{G} \right|_{k = m_{\text{Pl}}a} &=& 0,
\end{eqnarray}
for $k> m_{\text{Pl}}$.

\begin{figure}[ht!]
\centering
\includegraphics[width=8cm,angle=0]{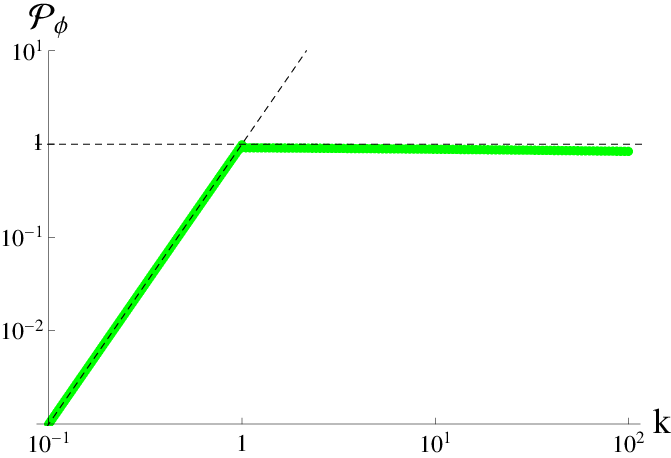}
\caption{Power spectrum  at $t=10$ $t_{\text{Pl}}$  with $w=-0.95$. The initial power spectrum was 
$\mathcal{P}(k)=k^3$ for $k<1$ and $\mathcal{P}(k)=1$ for the modes with $k>1$, imposed at the 
trans-Planckian surface, as depicted by the dashed lines. The slight red-tilt appears due 
to decrease of the Hubble factor in time. The values of $k$ are expressed in the Planck units.}
\label{PStot}
\end{figure}

Applying these initial condition to  Eqs. (\ref{EqP}) and (\ref{EqG}), the evolution of the power spectrum can be computed. 
The result is shown in Fig.  \ref{PStot}.
For $k<m_{\text{Pl}}$, the shape of the power spectrum is preserved because of the ``freezing" of modes at  
super-Hubble scales. The modes are initially super-Hubble and remain such during the 
whole evolution. The power spectrum for $k>m_{\text{Pl}}$ is slightly red-tilted due to 
gradual increase of  the Hubble radius during the evolution, as in usual cosmology. 

The spectrum is characterized by a sharp transition between the $k^3$ IR behavior and the nearly 
scale-invariant UV part. The sharpness is obviously due to the naive matching between 
initial conditions at the silence surface and at the Planckian surface but the existence of  two 
regimes is a specific prediction in this model. Some additional features are to be expected around 
$k\approx m_{\text{Pl}}$ with a more sophisticated modeling. 

\section{Flat power spectrum without trans-Planckian modes} 
\label{SuperPlanckFlat} 

In this section, we study if it is possible to convert a pure ``silent" initial power spectrum $\mathcal{P} \propto k^3$
to a flat one, in agreement with  observational data. To this aim, 
we will consider two successive periods dominated by different types of barotropic matter. 
Here, the quantum effects will not be neglected in the dynamics. They will be used to determine 
initial conditions. It has, however, been numerically checked that the subtle corrections to the 
propagation equations do not play a significant role in the shape of the power spectrum.

Initially, the Hubble horizon is of the order of the Planck scale, that means that all the relevant modes 
(those with wavelengths much bigger than the Planck length) are frozen: they are outside the Hubble 
radius, and therefore (approximately) constant in time. If the amplitude of a mode is to evolve, so that 
a final spectrum (called primordial for the subsequent phenomenology) compatible with observations 
emerges, one needs the modes to first enter the Hubble horizon and then to exit again. To achieve this, 
one needs a background where the conformal Hubble factor $\h=\frac{a'}{a}$ is first decreasing and then,
later on, growing. With barotropic matter, this means that at least two successive periods with different 
pressure to density ratios are required. Since entering and exiting the Hubble horizon is the key point 
here, and what happens to the background in-between is somehow irrelevant, there is no reason to 
include more than two different barotropic periods.

The important issue is to determine precisely the sufficient conditions  for this specific evolution 
to indeed generate a scale-invariant spectrum.  In order to answer this question let us 
consider the evolution of modes far from the surface of silence, where the $\Omega\approx 1$ approximation 
is valid. For a barotropic matter content, the solution to the mode equation (\ref{Ftensor}) is 
\begin{equation}
f_k = \frac{\sqrt{-k\eta}}{\sqrt{k}} \sqrt{\frac{\pi}{4}} \left(D_1 H_{|\nu|}^{(1)}(-k\eta)+D_2H_{|\nu|}^{(2)}(-k\eta)\right),
\label{solBarotropic}
\end{equation}
where 
\begin{equation}
|\nu| = \frac{3}{2} \left|\frac{1-w}{1+3w}\right|, 
\end{equation}
and $D_1$ and $D_2$ are constants which, based on the Wronskian condition, satisfy
the following relation
\begin{equation}
|D_1|^2-|D_2|^2=1. 
\end{equation}
Using the small value approximation of the Hankel function 
\begin{equation}
H_{|\nu|}^{(1)}(x) \simeq  -\frac{i}{\pi} \Gamma(|\nu|) \left( \frac{x}{2} \right)^{-|\nu|},
\end{equation}  
the power spectrum at the super-Hubble scales is 
\begin{equation}
\mathcal{P}_{\phi}(k) \propto \left|D_1-D_2\right|^2k^{3-2|\nu|}.  
\label{PowerBarotropic} 
\end{equation}
Assuming the Bunch-Davies vacuum at short scales ($D_1=1$, $D_2=0$), a flat power spectrum is obtained for $|\nu|=\frac{3}{2}$.
This, in the expanding universe, corresponds to $w=-1$.  In this case, the power spectrum at 
the short scales is 
\begin{equation}
\mathcal{P}_{\phi}(k) \propto k^2.    
\end{equation}
The quadratic sub-Hubble spectrum has to be produced from the initial cubic 
super-Hubble modes.\\

This can be achieved by considering a prior evolution driven by the barotropic matter. 
The evolution of the modes is described by Eq. (\ref{solBarotropic}). The 
initial cubic power spectrum at the super-Hubble scales can be obtained by taking 
$|\nu|=0$ (which corresponds to $w=1$) and requiring $D_1$ and $D_2$ to be independent on $k$. 
In that case, using the large value approximation,
\begin{equation}
H_{|\nu|}^{(1)}(x) \simeq   \sqrt{\frac{2}{\pi x}} e^{i(x-|\nu|\pi/2-\pi/4)},
\end{equation}
the sub-Hubble spectrum is 
\begin{equation}
\mathcal{P}_{\phi}(k) \propto k^2     
\end{equation}
as required.\\

\begin{figure}[ht!]
\centering
\includegraphics[width=8cm,angle=0]{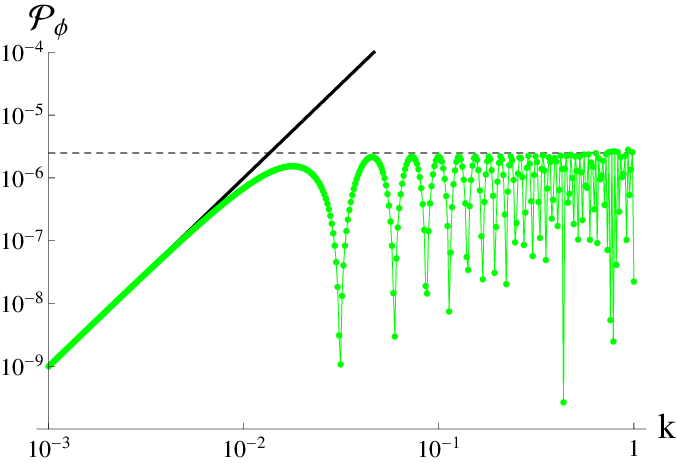}
\caption{Power spectrum converted to a rough scale invariance from the initial $\mathcal{P}(k)=(k/m_{\text{Pl}})^3$. 
The spectrum was computed at $t=6000 t_{\text{Pl}}$ 
with $\frac{\Lambda}{8\pi G \rho_c} = 10^{-7}$. } 
\label{ConSpect}
\end{figure}

The transition between the phases characterized by $w=1$ and $w=-1$ can be 
modeled by considering the following energy density:
\begin{equation}
\rho = \frac{\rho_0}{a^6}+\frac{\Lambda}{8\pi G}. 
\end{equation} 
The first factor corresponds to matter with an equation of state $P=\rho$ while the 
second is the cosmological constant term with the equation of state $P=-\rho$.
For such a setup together with initial conditions 
\begin{eqnarray}
\left. \mathcal{P} \right|_{\Omega=0} &=& (k/m_{\text{Pl}})^3,  \\
\left. \mathcal{G} \right|_{\Omega=0} &=& 0,
\end{eqnarray}
for $k< m_{\text{Pl}}$, the equations of motion can be solved. In Fig. \ref{ConSpect} we show 
the results of the numerical simulations with $\frac{\Lambda}{8\pi G \rho_c} = 10^{-7}$.

As expected, the spectrum scales as $k^3$ in the IR limit and becomes quasi-flat 
for  sufficiently large values of $k$. The flatness is however distorted by 
 acoustic oscillations resulting from the sub-Hubble evolution of modes.  

The above argument can be generalized. A cubic spectrum will be transformed to a 
flat spectrum under the following conditions: the relevant modes enters the Hubble 
horizon when $\omega=\omega_1>-1/3$ and exits the Hubble horizon when 
$\omega=\omega_2<-1/3$, where $\omega_1$ and $\omega_2$ fulfill
\be
\frac{1}{1+3\omega_1}-\frac{1}{1+3\omega_2}=\frac{3}{4}.
\ee
In any case agreeing with the above conditions, the final spectrum will be flat 
and modulated by acoustic oscillations.

\section{Conclusions} \label{Summ}

Loop quantum cosmology, together with other approaches to quantum gravity, might lead to a 
stage of asymptotic silence in the early Universe. This is in remarkable agreement with 
expectations from the purely classical BKL conjecture. In itself, the possible existence of 
this specific situation opens many windows to make links between primordial cosmology 
and phase transitions in solid state physics, not to mention the key question of  ``emergence" 
of time. In this work, we focused on more specific aspects, namely on the derivation of the 
tensor power spectrum of cosmological perturbations with initial conditions imposed 
around the state of silence. 

The main results of the paper are the following:
\begin{itemize}
\item The state of silence provides a  natural way to set initial conditions for the 
cosmological perturbations. This state is the beginning of the Lorentzian phase, 
and, in some sense, the beginning of time.  This is qualitatively a new feature in cosmology. 
\item A cubic shape ($\mathcal{P} \propto k^3$) of the initial power spectrum at the ``silent surface''
is favored. 
\item In the deep Euclidean regime, the state of vacuum can be defined for some 
phantom-like equation of states at large scales, which contrasts with the Lorentzian 
case. Such vacuum is generically characterized by a  $\mathcal{P} \propto k^3$ power spectrum (white noise).    
\item If the evolution of the universe was such that the modes of physical relevance were to become 
trans-Plackian when evolved backward in time up to the ``silent surface'', then 
initial conditions can be set at the ``Planckian surface'' (which is of course time dependent). 
The scale-invariant power spectrum of cosmological perturbations can be obtained if the power 
spectrum at the Planckian surface has a constant value. 
\item If the visible part of the power spectrum is due to modes that emerged from the silent 
surface without having been trans-Planckian, the white noise initial power spectrum can indeed 
be turned into a quasi-flat primordial power spectrum without inflation but this requires 
a quite high level of fine tuning. In this case the super-Hubble spectrum ($\mathcal{P} \propto k^3$)
is converted to a $\mathcal{P} \propto k^2$ spectrum at sub-Hubble scales and the modes cross 
the horizon once again leading to a``final" $\mathcal{P} \approx$ const spectrum. 
\end{itemize}

\section*{Acknowledgments}

The work was supported by Astro-PF and HECOLS programs, and by the Labex ENIGMASS.
JM is supported by the Grant DEC-2014/13/D/ST2/01895 of the National Centre of Science.

\end{document}